\documentclass[twocolumn,showpacs,preprintnumbers,amsmath,amssymb,prl]{revtex4}

\usepackage{graphicx}
\usepackage{dcolumn}
\usepackage{bm}

\newcommand{\beq}{\begin{equation}}
\newcommand{\eeq}{\end{equation}}
\newcommand{\beqar}{\begin{eqnarray}}
\newcommand{\eeqar}{\end{eqnarray}}
\newcommand{\bcen}{\begin{center}}
\newcommand{\ecen}{\end{center}}
\newcommand{\bitem}{\begin{itemize}}
\newcommand{\eitem}{\end{itemize}}

\newcommand{\wf}{\bm{\Psi}}
\newcommand{\cheb}{\bm{\Phi}}
\newcommand{\rar}{\rightarrow}

\begin{document}
\preprint{}

\title{
       A Pulse Shaping Algorithm of a Coherent Matter Wave \\
       Controlling Reaction Dynamics.
       }

\author{Solvejg J\o rgensen}
 \email{solvejg@fh.huji.ac.il}
\author{Ronnie Kosloff}%
 \email{ronnie@fh.huji.ac.il}
\affiliation{%
The Fritz Haber Research Center for Molecular Dynamics,\\
Hebrew University, Jerusalem 91904, Israel
}%

\date{\today}

\begin{abstract}
A pulse shaping algorithm for a matter wave with the purpose
of controlling a binary reaction has been designed.
The scheme  is illustrated for an Eley-Rideal reaction
where an impinging matter-wave atom 
recombines with an adsorbed atom on a metal surface.
The wave function of the impinging atom is shaped such that 
the desorbing molecule leaves the surface in a specific vibrational state.

\end{abstract}

\pacs{{82.53.Kp}, {03.75.-b}, {03.75.Pp}, {32.80Qk}, {34.50.Dy}}
\maketitle

A binary chemical reaction can be controlled by means of 
interference of matter waves. A source of coherent matter waves for such a control 
scheme could originate from an output coupler of a Bose-Einstein 
condensate\cite{MOMewes97,BPAnderson98,EWHagley99,IBloch99}.
A two-pulse scenario for such a control  has been previously explored  and termed (2PACC) 
spectroscopy\cite{2pacc:1,2pacc:2}. In the 2PACC model the control knobs 
were limited to the time delay between the pulses and the phase difference.  
It was demonstrated that the control was able to obtain a significant enhancement 
of the probability of creating a gas phase molecule in an Eley-Rideal reaction.
The drawback of the 2PACC scheme was that the desorbing molecule 
possessed a broad distribution  of vibrational states. 
The present study demonstrates that additional control
on the incoming matter wave can create a specific final vibrational state.

To achieve this target the task is to figure out how to shape the incoming wavepacket. 
An iterative computation scheme is developed for this task which is in analogy 
to   the 
optimal control theory of optical pulse shaping \cite{k67,SShi90,THornung01,JPPalao02}.
As important is the development of experimental schemes able to shape such a wavepacket.
 
As in earlier studies\cite{2pacc:1,2pacc:2}, the control target is the  Eley-Rideal
bimolecular reaction where an impinging atom recombines with an adsorbed 
atom on a metal surface to form a diatomic molecule 
\beq
{\bf Y} + {\bf H}/{\bf Cu}(111) \rar {\bf YH}  + {\bf Cu}(111)\;\;\;.
\eeq
The source of atom {\bf Y} is a matter wave, 
which in the presented model consists of either hydrogen or alkali atoms.
The matter-wave of {\bf Y} is directed to a {\bf Cu}(111)-surface
with low coverage chemisorbed hydrogen atoms.
When the wave function of {\bf Y} overlaps with the one of the adsorbed atom,
interaction is expected, leading to a recombination that forms
the {\bf YH} molecule. If the newly formed molecule has sufficient
kinetic energy it will eventually desorb from the surface.
By shaping the wave function, the vibrational state of the desorbing
molecules becomes controlled by constructive or destructive quantum interferences.

\begin{figure}[b!]
{\includegraphics[scale=0.4]{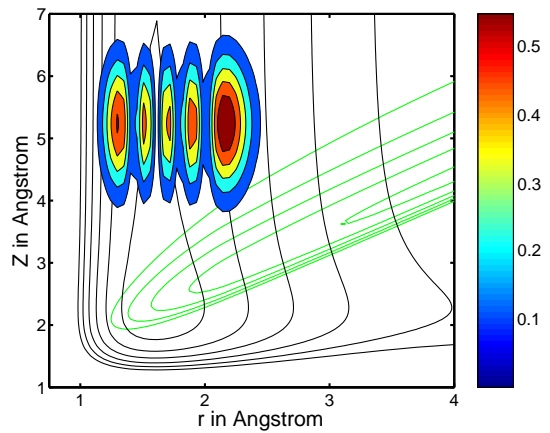}}
\caption{
\label{fig:initial} 
(Color)
An initial trial wave function is shown for the pulse shaping scheme
for the Eley-Rideal reaction where an impinging lithium recombines 
with chemisorbed hydrogen atom on a Cu(111)-surface.
This wave function is used for shaping the matter wave
which leads the molecule to desorbs in the fifth vibrational state ($v$$=$4).
}
\end{figure}
For a two electronic state case, the wave function of the system is described by a vector
\beq
\wf(r,Z,t)=\left ( \begin{array} {c}
                          \wf_{R}(r,Z,t) \\
                          \wf_{P}(r,Z,t)
                   \end{array}
            \right ) \;\;\;,
\eeq
where the reactant and product channels
are denoted by the index $\{R,P\}$.
The two degrees of freedom are the intramolecular distance
between the two atoms ($r$) and the distance from the surface to the
center-of-mass motion of the molecule ($Z$).

\begin{figure*}[t!]
{\includegraphics[scale=0.80]{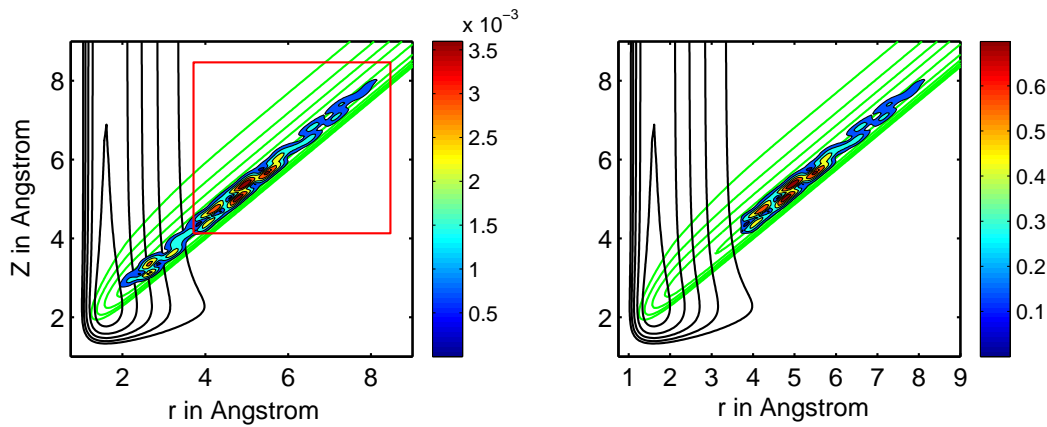}}
\caption{
\label{fig:pick}
(Color)
Above, the selection step of the pulse shaping scheme is illustrated.
A wave function on the reactant channel, ${\bm \phi}_R^f$,
is shown after termination of the
backward time propagation (left).
The wave packet of the red box has been selected to represent the 
wave packet of the matter wave on the reactant channel.
The renormalized selected wave function, ${\bm \phi}_R$,
which represents the wave function of the matter
wave is shown (right). This wave packet is used for the backward propagation.
}
\end{figure*}

\begin{figure*}[tb!]
{\includegraphics[scale=0.80]{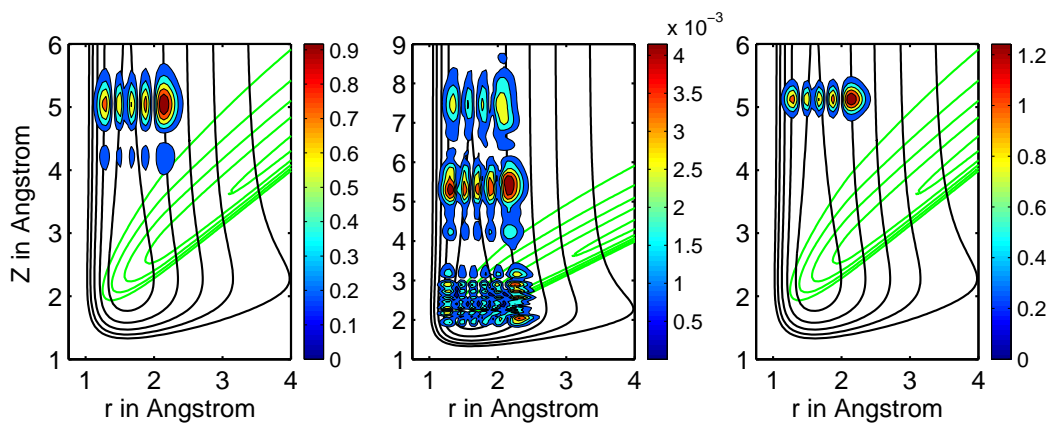}}
\caption{
\label{fig:proj}
(Color)
Above, the projection step of the pulse shaping scheme is illustrated.
The trial wave function before the projection is showed on the left.
The wave function on the product channel, ${\bm \phi}_R^b$, 
   after termination of forward time propagation (center).
The trial wave function after projection (right).
}
\end{figure*}

The dynamics of the Eley-Rideal reaction was followed by solving the
time-dependent two-channel Schr\"odinger equation.
The pulse shaping algorithm is based on propagating the wave function forward 
($\tau>0$) and backward ($\tau<0$) in time.
The Chebychev propagation method\cite{Kosloff94b} was employed since it allows 
the propagation of the wave function in both time directions
\beq
\wf(r,Z,t+\tau) = e^{ - i \omega \tau} 
                 \sum_{k=0}^{\infty} a_k (\nu \tau) \cheb_k(r,Z) \;.
\label{eq:firsttrialwf}
\eeq
The coefficients of the expansion, $a_k$, are proportional to 
the k'th Bessel functions $J_k$. 
These coefficients depend on the sign of the time propagation.
The time symmetry enters in the Bessel functions $J_k$,
the Bessel functions of odd order change sign as the direction of
time propagation changes.
The functions, $\cheb_k$, are obtained by operating $T_k({\bm H}_N)$
on $\wf(r,Z,t)$, where $T_k$ are the Chebychev polynomials.
The normalized Hamiltonian ${\bm H}_N$ is a shift-and-scaled version
of the original Hamiltonian, enforcing the eigenvalues of the Hamiltonian 
into the interval $[$$-$$1,1]$.
The parameters, $\omega$ and $\nu$, are a result of the normalization 
of the Hamiltonian.

\begin{figure*}[bt!]
{\includegraphics[scale=0.80]{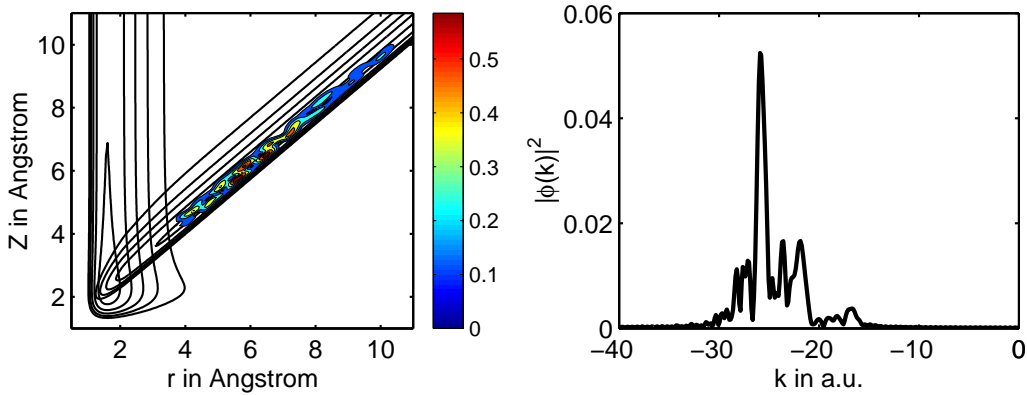}}
\caption{
\label{fig:wfatom}
(Color)
The wave function on the reactant channel which leads to 67 $\%$ of molecules
desorb in the fifth vibrational state (left)
Fourier transform of the a one-dimensional wave function represent the 
impinging atom for fixed distance between the surface and the chemisorbed
atom at 0.916{\AA} (right).
}
\end{figure*}
The task is to find the initial wavepacket on the reactive channel
which will lead to desorption of the molecule in the $v$'th vibrational state. 
The scheme is based on reversing the reaction e.g. starting from a
product wave function representing the desorbing molecule 
in the $v$'th vibrational state. This wave function is propagated backwards in time
to find  the reactive wave function leading to the desired final condition.
The scheme is illustrated for an Eley-Rideal reaction in which an impinging 
lithium atom recombines with an adsorbed hydrogen atom. 
A trial wave function on the product channel has been constructed.
Far enough from the surface the internal and external  degrees of freedom 
are uncoupled. Thus  the trial wave packet can be expressed 
as a product of the $v$'th molecular  vibrational eigenstate, 
$\chi_v (r)$ ($v$$=$$\{0,1,...\}$), and a Gaussian wavefunction, $g(Z)$, 
\beq
\wf_{trial}(r,Z)=\left ( \begin{array} {c}
                             {\bm 0}             \\
                             \chi_n (r) g(Z)
                         \end{array}
                  \right ) \;\;\;.
\eeq
The wavefunction used for control is the wavefunction of a lithium matter wave 
source. The target is to desorb a lithium hydride molecule  
in the fifth vibrational state ($v=4$).
A target trial wave function has been constructed 
according to Eq.\ref{eq:firsttrialwf} and 
is shown in Fig. \ref{fig:initial}.
An iterative scheme is used for inversion i.e.
finding  the condition for the wavepacket on the reactant
channel leading  to desorption of the molecules in 
the $v$'th vibrational state.
\begin{enumerate}

\item {\em {Backward propagation:}}
      A trial wave function $\wf_{trial}(r,Z)$ is propagated backward in time.
      As the molecule approaches the surface 
      population is transferred to the reactant channel through
      the non-adiabatic coupling. 
      The part of the wave packet which is transferred to the reactant channel
      corresponds to a molecular dissociation. That is one atom is chemisorbed
      on surface and the other desorbing to the gas phase.
      After a preassigned time the propagation is stopped, 
      leading to the wave function $\wf_1=({\bm \phi}_R^f,{\bm \phi}_P^f)$.
      Fig. \ref{fig:pick} (left) shows a wave packet in the reactant channel
      after termination of the backward time propagation.

\item {\em {Selection:}}
      Part of the backward propagating 
      wave function, $\wf_1$ is selected consisting of a
      wavefunction outside the interaction region.
      In Fig. \ref{fig:pick} (left) the wave function is shown 
      before selection. 
      The part of the wave function which is inside the red box has 
      been selected to resemble the wave function of the reactive channel. 
      The selected wave function is renormalized and 
      it is shown in Fig. \ref{fig:pick} (right).
      The size and location of the box is chosen such that the energy 
      of the selected and normalized wavepacket is close to the energy
      of the trial wavepacket.

\item {\em {Forward propagation:}}
      The selected wavefunction, $\wf_2=({\bm \phi}_R,{\bm 0})$, 
      is propagated forward in time. 
      Through the non-adiabatic coupling between two potential energy surfaces.
      The newly formed  molecules desorbs leading to the wavefunction
      $\wf_3=({\bm \phi_R^b},{\bm \phi_P^b})$.

\item {\em {Projection:}}
      The overlap between the forward propagated wave function, $\wf_3$,
      and the trial wavefunction is evaluated as a  function of 
      the center of mass
      \beq
      c_n(Z)=\int \wf_{trial}(r,Z)^* \wf_3 (r,Z) dr \;\;.
      \eeq
      A new trial wavefunction is created
      \beq
      \wf_{proj}(r,Z)=\left ( \begin{array} {c}
                         {\bm 0}             \\
                        c_n(Z) f(Z) \chi_n(r)
                     \end{array}
              \right ) \;\;\;.
      \eeq
      In the first iteration $f(Z)$ is $g(Z)$.
      This new wave function is renormalized and it is used as 
      a new trial wave function in step 1 $\wf_{trial}=\wf_{proj}$.
      The projection of the forward propagated 
      wave packet is illustrated in Fig. \ref{fig:proj}.
\end{enumerate}

\begin{figure}[tb!]
{\includegraphics[scale=0.4]{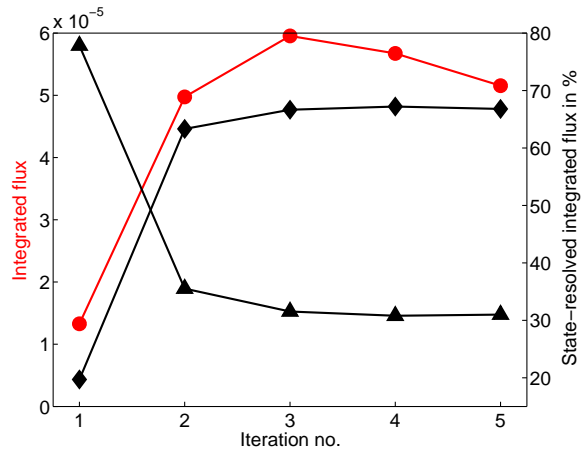}}
\caption{
\label{fig:iter} 
(Color)
The accumulated flux (red line) is shown as a function of the iteration number.
The percentage of the accumulated flux in the forth and fifth state have 
been computed as a function of the iteration number.
The forth and fifth vibrational states are shown with triangles and diamond,
respectively.
}
\end{figure}

The above pulse shaping scheme is repeated a number of times.
It is important that the energy of the total system is conserved
after either the selection stage  or projection of the total wave function.
The pulse shaping algorithm is stopped when a satisfactory yield 
in the desired vibrational state has been reached.
The ultimate goal is that all the flux of desorbing molecules 
is probed into the particular vibrational state.
In order to improve the convergence of the algorithm the number of grid points
has been enlarged (1024$\times$1024) and the grid spacing has been reduced 
($dr$$=$$dZ$$=$0.05a.u.) compared to the earlier 2PACC calculations\cite{2pacc:2}.
The potential energy surfaces are given in Ref.\cite{2pacc:2}, 
where the non-adiabatic coupling term was increased to $\beta$$=$0.1eV.
The backward propagation is stopped 
when the non-adiabatic population transfer is negligible.
The size and location of the box used for selecting the wave packet 
in step 2 vary from iteration to iteration in order to keep 
the reactive wavepacket inside the box. 
The forward propagation is terminated 
when the overlap between the trial wavefunction
and the propagated wavefunction is maximized. 

Starting from the trial wavefunction of Fig. \ref{fig:initial} 
the pulse shaping scheme was used to
obtain the optimal wavepacket in the reactant channel leading to desorption
of the molecule in the fifth vibrational state.
Fig. \ref{fig:iter} shows the accumulated flux of desorbing 
molecules as a function of the iteration number.
State-resolving the accumulated flux enables to calculate the percentage of desorbing 
molecules in the fourth and fifth vibrational states Cf. Fig.  \ref{fig:iter}.
A monotonic increase of the yield in $v=4$ is observed.

The optimal reactant wavefunction is shown in  Fig. \ref{fig:wfatom} (left). 
The wave function of the matter wave is projected out for 
the equilibrium distance between the chemisorbed atom 
and the surface (0.916{\AA}).
The 1D wave packet is found to be multi-pulsed reconfirming  previous studies, 
that control is obtained by destructive or constructive quantum interferences. 
The momentum representation is obtained by
performing a Fourier transform of the one-dimension matter-wave.
The momentum components are shown in Fig. \ref{fig:wfatom}. 
Several momentum components can be observed which are needed for creating 
the necessary constructive quantum interference in the fifth vibrational state 
and destructive in all the other vibrational states.

The source of the matter-waves, the BEC condensate,
has to be positioned very close to a surface.
Such a device is realized in the so-called
atom chips\cite{SSchneider03,MPAJones03,RFolman02,WHansel01}
or surface micro-traps\cite{AELeanhardt02,HOtt01}.
In the atom chips for example the BEC has been placed a few hundred microns
above a metal surface\cite{RFolman02}. 

The experimental realization of the control scheme depends on the 
the ability to shape matter-waves. Four-wave mixing\cite{LDeng99}
of matter waves has been demonstrated which can lead to 
different momentum components of the matter-waves. 
A more comprehensive solution could follow ideas from optical pulse shaping.
A matter-wave diffraction grid has been demonstrated\cite{KHornberger03,RBDoak99}. 
A setup of
two diffraction grids and a phase shift device would constitute the ultimate
matter-wave pulse shaper. Such a device would make the suggested control scheme
realizable.

SJ thanks Marie Curie Fellowship Organization.
This work was supported by the Israel Science Foundation
and the European Research and Training Network
{\em Cold Molecules: Formation, Trapping and Dynamics}.
The Fritz Haber Center is supported by the Minerva
Gesellschaft f\"ur die Forschung, GmbH M\"unchen, Germany.

\end{document}